\documentclass[letter,longauth]{aa}  
\usepackage{epsfig}
\usepackage{natbib}
\usepackage{txfonts}
\usepackage{graphicx}

\def\ms{\hbox{\,m/s}}         
\def\m2s2{\hbox{\,m$^{2}$\,s$^{-2}$}} 
\def\kms{\hbox{\,km/s}}       
\def\Msun{\hbox{$M_{\odot}$}}             
\def\Rsun{\hbox{$R_{\odot}$}}
\def\Mjup{\hbox{$M_{\rm Jup}$}}
\def\Rjup{\hbox{$R_{\rm Jup}$}}
\def\degr{\hbox{$^\circ$}}

\def\chisq{\mbox{$\chi^2$}}

\begin{document}

\title{Transiting exoplanets from the CoRoT space mission}
\subtitle{II. CoRoT-Exo-2b: A transiting planet around an active G star\thanks{Based on observations obtained with CoRoT, a space project operated by the French Space Agency, CNES, with participation of the Science Programme of ESA, ESTEC/RSSD, Austria, Belgium, Brazil, Germany and Spain; and on observations made with SOPHIE spectrograph at Observatoire de Haute Provence, France (PNP.07 A.MOUT), CORALIE, and HARPS spectrograph at ESO La Silla Observatroy (079.C-0127/F)) } }

\author{Alonso, R. \inst{1}
\and Auvergne, M. \inst{2}
\and Baglin, A. \inst{2}
\and Ollivier, M. \inst{3}
\and Moutou, C. \inst{1}
\and Rouan, D. \inst{2}
\and Deeg, H.J. \inst{4}
\and Aigrain, S. \inst{5}
\and Almenara, J.M. \inst{4}
\and Barbieri, M. \inst{1}
\and Barge, P. \inst{1}
\and Benz, W. \inst{6}
\and Bord\'e, P. \inst{3}
\and Bouchy, F. \inst{7}
\and De la Reza, R. \inst{8}
\and Deleuil, M. \inst{1}
\and Dvorak, R. \inst{9}
\and Erikson, A. \inst{10}
\and Fridlund, M. \inst{11}
\and Gillon, M. \inst{12}
\and Gondoin, P. \inst{11}
\and Guillot, T. \inst{13}
\and Hatzes, A. \inst{14}
\and H\'ebrard, G. \inst{7}
\and Kabath, P. \inst{10}
\and Jorda, L. \inst{1}
\and Lammer, H. \inst{15}
\and L\'eger, A. \inst{3}
\and Llebaria, A. \inst{1}
\and Loeillet, B. \inst{1,7}
\and Magain, P. \inst{16}
\and Mayor, M. \inst{12}
\and Mazeh, T. \inst{17}
\and P\"atzold, M. \inst{18}
\and Pepe, F. \inst{12}
\and Pont, F. \inst{12}
\and Queloz, D. \inst{12}
\and Rauer, H. \inst{9,19}
\and Shporer, A. \inst{17}
\and Schneider, J. \inst{20}
\and Stecklum, B. \inst{14}
\and Udry, S. \inst{12}
\and Wuchterl, G. \inst{14}
}

\offprints{\email{roi.alonso@oamp.fr}}

\institute{Laboratoire d'Astrophysique de Marseille, UMR 6110, CNRS/Universit\'e de Provence, Traverse du Siphon, 13376 Marseille, France
\and
LESIA, CNRS UMR 8109, Observatoire de Paris, 5 place J. Janssen, 92195 Meudon, France
\and
IAS, Universit\'e Paris XI, 91405 Orsay, France
\and
Instituto de Astrof\'\i sica de Canarias, E-38205 La Laguna, Spain
\and
School of Physics, University of Exeter, Stocker Road, Exeter EX4 4QL, United Kingdom
\and
Physikalisches Institut, University of Bern, Sidlerstrasse 5, 3012 Bern, Switzerland
\and
Institut d'Astrophysique de Paris, UMR7095 CNRS, Universit\'e Pierre \& Marie Curie, 98bis Bd Arago, 75014 Paris, France
\and
Observat\'orio Nacional, Rio de Janeiro, RJ, Brazil
\and
Institute for Astronomy, University of Vienna, T\"urkenschanzstrasse 17, 1180 Vienna, Austria
\and
Institute of Planetary Research, DLR, Rutherfordstr. 2, 12489 Berlin, Germany
\and
Research and Scientific Support Department, European Space Agency, ESTEC, 2200 Noordwijk, The Netherlands 
\and
Observatoire de Gen\`eve, Universit\'e de Gen\`eve, 51 Ch. des Maillettes, 1290 Sauverny, Switzerland
\and
Observatoire de la C\^ote d'Azur, Laboratoire Cassiop\'ee, CNRS UMR 6202, BP 4229, 06304 Nice Cedex 4, France
\and
Th\"uringer Landessternwarte Tautenburg, Sternwarte 5, 07778 Tautenburg, Germany
\and
Space Research Institute, Austrian Academy of Sciences, Schmiedlstrasse 6, 8042 Graz, Austria
\and
Institut d'Astrophysique et de G\'eophysique, Universit\'e de  Li\`ege, All\'ee du 6 ao\^ut 17, Sart Tilman, Li\`ege 1, Belgium
\and
School of Physics and Astronomy, R. and B. Sackler Faculty of Exact Sciences, Tel Aviv University, Tel Aviv 69978, Israel
\and 
Rheinisches Institut f\"ur Umweltforschung, Universit\"at zu K\"oln, Abt. Planetenforschung, Aachener Str. 209, 50931 K\"oln, Germany
\and
Center for Astronomy and Astrophysics, TU Berlin, Hardenbergstr. 36, D-10623 Berlin, Germany
\and
LUTH, Observatoire de Paris-Meudon, 5 place J. Janssen, 92195 Meudon, France
}

\date{Received / Accepted }

\abstract 
{The CoRoT mission, a pioneer in exoplanet searches from space, has completed its first 150~days of continuous observations of $\sim$12\,000 stars in the galactic plane. An analysis of the raw data identifies the most promising candidates and triggers the ground-based follow-up.} 
{We report on the discovery of the transiting planet CoRoT-Exo-2b, with a period of 1.743~days, and characterize its main parameters.}
{We filter the CoRoT raw light curve of cosmic impacts, orbital residuals, and low frequency signals from the star. The folded light curve of 78 transits is fitted to a model to obtain the main parameters. Radial velocity data obtained with the SOPHIE, CORALIE and HARPS spectrographs are combined to characterize the system. The 2.5~min binned phase-folded light curve is affected by the effect of sucessive occultations of stellar active regions by the planet, and the dispersion in the out of transit part reaches a level of 1.09$\times$10$^{-4}$ in flux units.}
{We derive a radius for the planet of 1.465$\pm$0.029~\Rjup \,and a mass of 3.31$\pm$0.16~\Mjup, corresponding to a density of 1.31$\pm$0.04~g/cm$^3$. The large radius of CoRoT-Exo-2b cannot be explained by current models of evolution of irradiated planets.}
{}


\titlerunning{CoRoT-Exo-2b}

\authorrunning{Alonso et al.}

\maketitle

\section{Introduction}
\label{sec:intro}
The search for transiting extrasolar planets is entering a new era with the first dedicated space-based project CoRoT \citep{2006cosp...36.3749B}. Launched in December 2006, it continuously monitors around 12\,000 stars per observing run, with durations of $\sim$30~days for the \emph{short runs} and $\sim$150~days for the \emph{long runs}. CoRoT circumvents the main limitations of ground-based transit searches, i.e., the effects of the Earth atmosphere and the reduced observing duty cycle. Since February 2007, CoRoT has been producing high quality light curves, whose first analysis is very promising (Auvergne et al.,~\emph{in preparation}). \cite{barge08} reported on the discovery of the low density transiting planet CoRoT-Exo-1b, the first transiting planet detected from space, during the first observing run of the mission.

CoRoT-Exo-2b is the second confirmed transiting planet discovered by CoRoT. In this work, we report on its early detection by the application of the \emph{alarm mode} to the raw data, and estimate its orbital and physical parameters from an analysis of the CoRoT phase-folded light curve and the radial velocity observations taken with different spectrographs.
\section{Observations}
\label{sec:obs}
The first \emph{long run} of CoRoT, which pointed towards the galactic center direction, started on May 16th, 2007 and ended on October 15th, constituting of a total of 152 days of almost-continuous observations. CoRoT-Exo-2b was identified as a planetary candidate by the alarm mode, which is aimed at identifying a reduced number of stars (500 out of 6000 in each of the 2 CCDs) for which the sampling rate is changed from 512~sec to 32~sec. For the first long run, this operational mode of CoRoT worked with raw data in ``white'' light, to minimize the time delay between the observations and the detection of the transit planet candidates. The quality of these data is such that we were able to easily detect Jupiter-sized transiting objects, and trigger follow-up observations while CoRoT was still acquiring data. This was the case for CoRoT-Exo-2, whose radial velocity and ground-based photometric follow-up started in July 2007. 

The star showed an apparent close companion (4\arcsec) to the main target, 3.5~mags fainter in $V$. Follow-up photometry from different sites (Wise Observatory 1~m telescope in Israel, IAC-80~cm telescope in Spain, and TLS-2~m in Germany), with higher spatial resolution than the CoRoT masks, verified that the transit signature was produced on the brightest star, thus minimizing the possibility of a confusion with a background eclipsing binary. An examination of the BEST \citep{2004PASP..116...38R} archived observations on the CoRoT fields showed a transit event in July 2005 that served to refine the early ephemeris used to plan the different follow-up efforts.
\section{Analysis of the CoRoT light curve}
\label{sec:ana}
\begin{figure*}[!th]
\begin{center}
\epsfig{file=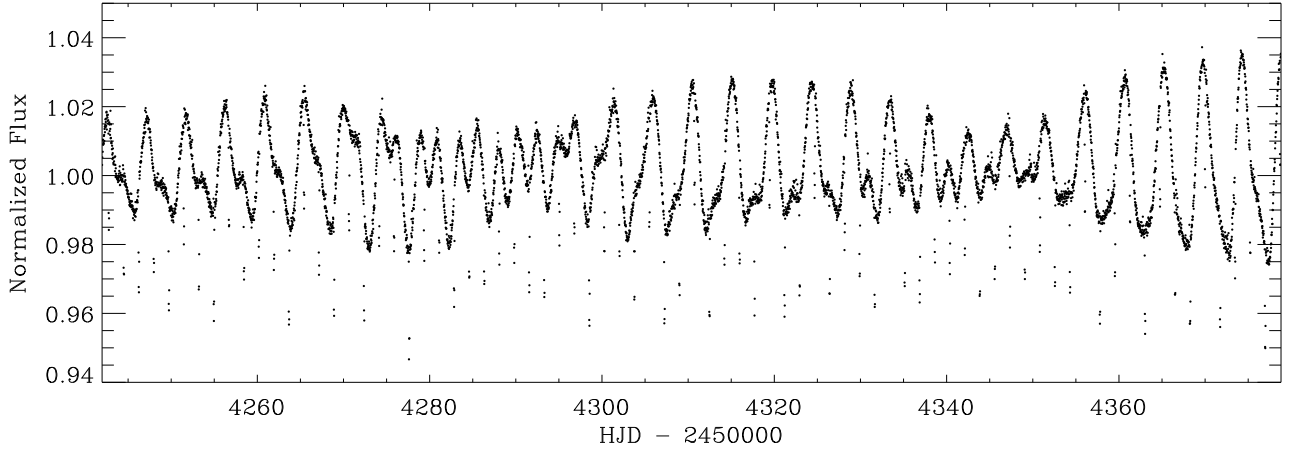,width=18cm}
\caption{Normalized flux of the CoRoT-Exo-2 star, showing a low frequency modulation due to the presence of spots on the stellar surface, and the 78 transits used to build the phase-folded transit of the Figure~\ref{fig:fig2}. For clarity purposes, data have been combined in 64-points bins ($\sim$34~min).} 
\label{fig:fig1}
\end{center}
\end{figure*}
The alarm data of CoRoT-Exo-2b consist of $\sim$369000 flux measurements with a sampling of 512~sec for the first week of data, and 32~sec for the rest of the run. These data were corrected for the CCD zero offset and gain. The contribution of the background light was estimated and corrected with a window containing 100~pixel and located close to the target.

To detect and eliminate the outlier data points, caused mainly by the crossing of the SAA (South Atlantic Anomaly), we subtracted a moving-median filtered version of the light curve and discarded the points at distances greater than 3.3-times the dispersion of the residuals. This way, 6.2\% of the data were rejected, achieving a duty cycle of $\sim$93.6\%.

A faint orbital signal (with a period of 1.7~hours) is expected due to the rough correction applied to the data. We calculated this remaining signal by folding the light curve with the orbital period between -25 and 25 orbits around each $j$-th orbit of the satellite, and by applying a median filter to this folded light curve. The resulting signal is then subtracted from the portion of the light curve acquired during the $j$-th orbit.  

The star shows a close neighbour that falls completely inside the CoRoT mask. To estimate the fraction of flux that comes from this star, we used the magnitudes from the EXODAT database of CoRoT sample stars (Deleuil et al., \emph{in preparation}), the modelled spectral distributions of \cite{1985ApJS...59...33P}, and the filter responses of the different bandpasses used to obtain the EXODAT magnitudes. With these ingredients, we constructed a model of two stars that reproduces the observed differences in colors. Interestingly, this scaled model is consistent with a late-K or early-M type companion star ($J$-$K$=0.84) located at the same distance as CoRoT-Exo-2, and thus possibly gravitationally bound, as has been found for other transiting planets \citep{bakos}. We integrated the scaled model into the CoRoT response function, and obtained a fraction of flux of 5.6$\pm$0.3\%. This fraction of the median value of the light curve was subtracted from the data, and the light curve was finally normalized by its median value, which is of 711000 ($e^-$/32s). 

The final light curve (Fig.~\ref{fig:fig1}) shows a total of 78 transits, embedded in a flux that exhibits periodic variations of the order of a few percent at several periods between 4.5 and 5 days, due to the presence of spots on the stellar surface. An analysis and interpretation of these variations is out of the scope of this paper, and will be presented elsewhere. To minimize the effect of this low frequency modulation on the estimation of the transit parameters, we performed a parabolic fit to the regions before and after each transit, and corrected the transit and its neighbourhood for this slope. We note that a simple linear fit did not properly remove the stellar variations, while higher order polynomials up to order 5 gave results consistent with those reported below. We estimated the ephemeris from a linear fit to the measured times of transit centers. Finally, we folded the data using this ephemeris, eliminated a few outliers (1\% of the remaining data points) with a procedure similar to that explained above, and binned it with a bin size of 0.001 in phase, corresponding to $\sim$2.5~min. The error bars on each bin were estimated as the standard deviation of the $N$-points inside the bin divided by the square root of $N$. The average 1-sigma error bar outside of the transits is 7$\times$10$^{-5}$ (the expected precision for photon noise limited data, using this bin size and the median value of the flux reported above is 6$\times$10$^{-5}$), rising to around 1.2$\times$10$^{-4}$ inside the transits. This larger dispersion inside the folded transit is due to the presence of inhomogeneties (spots) on the star surface, as has been noted for other transiting planets around active stars (\citealt{2007prpl.conf..701C} for TrES-1, \citealt{2007A&A...476.1347P} for HD189733). As the rotation period of the star and the orbital period of the planet are different, the signal of the many occultations of spots by the planet is averaged in the folded light curve. 

\begin{figure}
\begin{center}
\epsfig{file=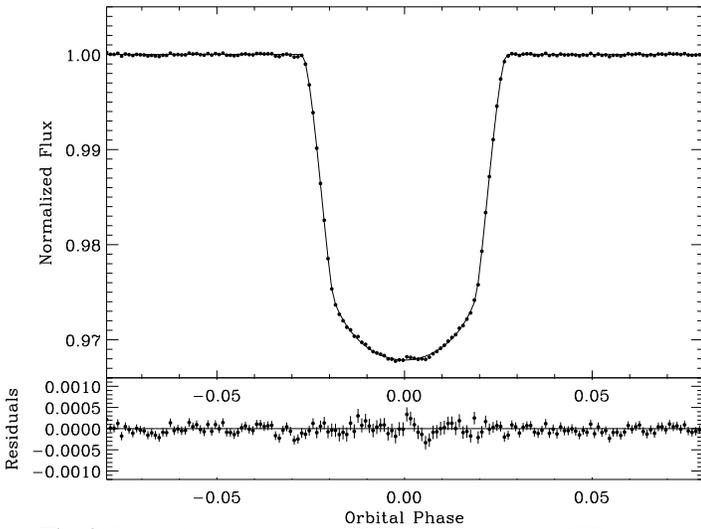,width=9cm,angle=0}
\caption{Normalized and phase folded light curve of 78 transits of CoRoT-Exo-2b (top), and the residuals from the best-fit model (bottom). The bin size corresponds to 2.5~min, and the 1-sigma error bars have been estimated from the dispersion of the points inside each bin. The residuals of the in-transit points are larger due to the effect of successive planet occultations of stellar active regions.} 
\label{fig:fig2}
\end{center}
\end{figure}

The transit was fitted to a model using the formalism of \citet{2006A&A...450.1231G}. To find the solution that best matches our data, we minimized the \chisq \, using the algorithm AMOEBA \citep{1992nrfa.book.....P}. The fitted parameters were the center of the transit, the phase of start of the transit $\theta_1$, the planet to star radius ratio $k$, the orbital inclination $i$ and the two non-linear limb darkening coefficients\footnote{We used a quadratic law for the limb darkenning, given by $I(\mu)=I(1)[1-u_a(1-\mu)-u_b(1-\mu)^2]$, where I is the distribution of brightness over the star and $\mu$ is the cosine of the angle between the normal to the local stellar surface and the line of sight. The use of $u_{+}$ and $u_{-}$ is a better choice to avoid correlations between the two limb darkening coefficients $u_a$ and $u_b$, as described in \cite{2006A&A...450.1231G}.} $u_{+}=u_a+u_b$ and $u_{-}=u_a-u_b$. To estimate the errors in each of the parameters we performed a bootstrap analysis with several thousand different sets of data, allowing for variations on the initial fitting parameters and taking into account the uncertainty on the fraction of the flux that comes from the close companion. To build each set, we: 1) subtracted the best solution to the data, 2) re-sorted a fraction of the residuals (1/$e$$\sim$37\%) and 3) added the subtracted solution in 1) to the new residuals. The errors were then estimated as the standard deviation from the fitted parameters. We also considered the effect of correlated flux residuals, by allowing the data to move as a whole by 2$\times$10$^{-4}$ in flux and repeating the fits, following the approach of \cite{2007A&A...476.1347P}. The results, reported in Table~\ref{tab:parpl}, show a compromise between the two methods. The phase folded light curve, the best fitted solution and the residuals around the fit are shown in Fig.~\ref{fig:fig2}.

The standard deviation of the residuals outside the transit phase is 1.09$\times$10$^{-4}$, which indicates the presence of uncorrected noise at a level of a few 10$^{-5}$. This noise is expected to be further reduced in future improvements to the CoRoT pipeline, in particular by the inclusion of a correction for the satellite jitter.
\begin{table}
\begin{minipage}[t]{\columnwidth}
\begin{center}
\caption[]{Parameters of the CoRoT-Exo-2 system. One-sigma errors are given, when relevant, in the last column. a) From \cite{bou08}. b) Zero albedo equilibrium temperature}  
\renewcommand{\footnoterule}{}  
\begin{tabular}{lrr}
\hline\hline
Star CorotID &\multicolumn{2}{c}{0101206560}\\
$RA$ (J2000) &\multicolumn{2}{c} {19$^h$27$^m$06.5$^s$}\\
$Dec$ (J2000) &\multicolumn{2}{c} {1$\degr$23$\arcmin$01.5$\arcsec$}\\
$V mag$& \multicolumn{2}{c}{12.57} \\
\hline
\multicolumn{3}{c}{\emph{Obtained from the CoRoT photometry}} \\
\hline
& Value& Error\\
$P$ [d]&  1.7429964&0.0000017\\
$T_{c}$ [BJD] &  2454237.53562& 0.00014\\
\multicolumn{2}{c}{\emph{Fitted}}&\\
$\theta_{1}$ & 0.02715&0.00008\\
$k=R_p/R_s$ &  0.1667&0.0006\\
$i$ [deg]& 87.84&0.10\\
$u_+$ & 0.471&0.019\\
$u_-$ & 0.34&0.04\\
\multicolumn{2}{c}{\emph{Deduced}}& \\
$u_a$ & 0.41&0.03\\
$u_b$ & 0.06&0.03\\
$a/R_{s}$ & 6.70&0.03 \\
$a/R_{p}$ & 40.3&0.3 \\
$M_{s}^{(1/3)}/R_{s}$  & 1.099&0.005 \\
\hline
\multicolumn{3}{c}{\emph{Photometry, spectroscopy and radial velocity combined}}\\
\hline
$V_0$ [\kms] & 23.245&0.010\\
$K$ [\kms] &  0.563&0.014\\
$e$ &  0 &(fixed)\\
$M_s$ [\Msun] & 0.97&0.06\\
$R_s$ [\Rsun] & 0.902&0.018\\
$v\sin i$ [km/s]$^a$ & 11.85& 0.50\\
$T_{eff}$ [K]$^a$ & 5625 & 120 \\
$M_p$ [\Mjup]  & 3.31&0.16\\
$R_p$ [\Rjup] & 1.465&0.029\\
$\rho_p$ [g/cm$^3$] & 1.31&0.04\\
$T_{eq}$ [K] $^b$& 1537 & 35\\
\hline
\label{tab:parpl}
\end{tabular}
\end{center}
\end{minipage}
\end{table}
\section{Radial velocities}
\label{sec:radvel}
Radial velocity observations of CoRoT-Exo-2 were performed in July 2007, at the 193cm telescope of the Observatoire de Haute Provence, France, with the SOPHIE spectrograph \citep{2006tafp.conf..319B}, and from La Silla Observatory (Chile) using both the 1.2~m Swiss telescope with the CORALIE spectrograph \citep{2000A&A...354...99Q}, and the 3.6~m telescope with the HARPS spectrograph \citep{2003Msngr.114...20M}.

\begin{figure}[!t]
\begin{center}
\epsfig{file=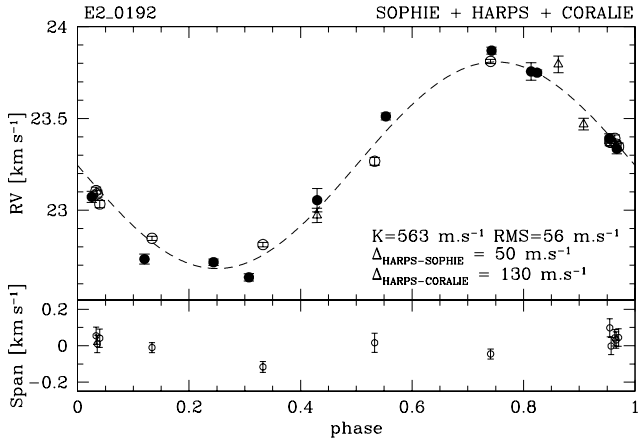,width=8.5cm}
\caption{Phase folded radial velocity measurements of CoRoT-Exo-2, together with the final fitted semi-amplitude ($K$) and the applied offsets between the instruments. Filled circles: SOPHIE, open circles: HARPS, open triangles: CORALIE. In the bottom panel, the total span of the CCF bisectors, as measured in the HARPS spectra. } 
\label{fig:fig3}
\end{center}
\end{figure}

Data reduction was performed with similar pipelines for the three spectrographs. A cross-correlation function (CCF) with a mask corresponding to a G2 star was calculated. The mean position, width, contrast, and bisector span of the cross-correlation function were then measured. A correction for Earth motion was also applied. In total, 11 measurements of the system were performed with SOPHIE, 3 with CORALIE and 12 with HARPS, spanning 83 days. Knowing the planet ephemeris, we could optimize the epoch of the measurements in order to cover all phases. In some measurements, the Moon signature in the cross-correlation function was identified, without a need to correct its contribution since the velocities were distant enough (more than 30~\kms). The radial velocities are given in Table~\ref{tab:obslog}, and the phase folded radial velocity measurements are plotted in Figure~\ref{fig:fig3}. 
%
\onltab{2}{
\begin{table}
\begin{center}
\caption[]{Radial velocity measurements of CoRoT-Exo-2 obtained by SOPHIE (S), CORALIE (C) and HARPS (H). The radial velocity shifts between the different spectrographs are not included.}  
\begin{tabular}{lllll}
\hline\hline
\small
BJD &  RV & Uncertainty & Instr.\\
 - 2400000 &      [km~s$^{-1}$] & [km~s$^{-1}$]& \\
\hline
  54295.48408 &   22.667 & 0.017 &S\\
  54296.49214 &   23.700 & 0.015 &S\\
  54298.46418 &   23.341 & 0.026 &S\\
  54298.48626 &   23.285 & 0.027 &S\\
  54298.58797 &   23.023 & 0.030 &S\\
  54303.44629 &   23.706 & 0.048 &S\\
  54304.51991 &   23.005 & 0.063 &S\\
  54308.55172 &   23.820 & 0.019 &S\\
  54309.53356 &   22.585 & 0.020 &S\\
  54313.45013 &   23.461 & 0.019 &S\\
  54314.44185 &   22.684 & 0.027 &S\\
  54329.67560 &  23.665 & 0.045 &C\\
  54330.66442 &  22.842 & 0.039 &C\\
  54378.56018 &  23.339 & 0.032 &C\\
  54345.52251 &  23.371 & 0.020 &H\\
  54345.52981 &  23.371 & 0.019 &H\\
  54345.53713 &  23.392 & 0.018 &H\\
  54345.54444 &  23.360 & 0.018 &H\\
  54345.55176 &  23.347 & 0.019 &H\\
  54345.65632 &  23.107 & 0.019 &H\\
  54345.66364 &  23.090 & 0.019 &H\\
  54345.67095 &  23.034 & 0.020 &H\\
  54346.53194 &  23.267 & 0.021 &H\\
  54347.57772 &  22.847 & 0.011 &H\\
  54348.63260 &  23.811 & 0.011 &H\\
  54349.66605 &  22.813 & 0.012 &H\\
\hline
\label{tab:obslog}
\end{tabular}
\end{center}
\end{table}
}
The radial-velocity points obtained show a variation in phase with the ephemeris derived from the CoRoT lightcurve. To fit these measurements, we applied a radial velocity shift between the different spectrographs (the values are displayed in Figure~\ref{fig:fig3}). The epoch and period of the transit were then fixed to the CoRoT values. Due to the very short period, we first assumed zero eccentricity. The semi-amplitude of the radial velocity variation and the mean velocity were then adjusted to the data. We repeated the fit with a free eccentricity, resulting in an orbit compatible with zero eccentricity (e$=$0.03$\pm$0.03). The final solution is displayed in Table \ref{tab:parpl}; the observed minus computed (O$-$C) residuals have a standard deviation of 56~\ms. This is significantly larger than the noise on individual measurements, and consistent with the expected effect of stellar activity, as described in \cite{bou08}. The semi-amplitude of the radial motion is large (K$=$563~\ms) due to the large planetary mass and very short period. These measurements thus establish the planetary nature of the transiting body detected by CoRoT and reject other interpretations such as a grazing or background eclipsing binary, or a triple system. The bisector of the CCF, plotted in Figure~\ref{fig:fig3}, shows no correlation of the spectral line shapes with the orbital period. Finally, the observation of the Rossiter-McLaughlin effect by  \cite{bou08} confirms the planetary nature of CoRoT-Exo-2b, as no triple system or blend could reproduce such a well-identified radial-velocity anomaly.
\section{Stellar and planetary parameters}
\label{sec:star}
From the fit to the folded light curve, we can measure with high precision the ratio $M_s^{1/3}/R_s$, and thus obtain a good estimation of the stellar radius once an estimation of its mass is given. The uncertainty on the stellar mass determination thus limits our knowledge of the stellar and planetary radii (as was the case in other space-based observations of transits, such as \citealt{2001ApJ...552..699B} or \citealt{2007A&A...476.1347P}). Assuming a mass of 0.97$\pm$0.06$M_\odot$ (Bouchy et al. 2008), we obtain a stellar radius of 0.902$\pm$0.018 $R_\odot$. This translates into a planetary mass of 3.31$\pm$0.16 \Mjup, a planetary radius of 1.465$\pm$0.029\Rjup \,and thus a planetary mean density of 1.31$\pm$0.04 g/cm$^3$.

Using models of the evolution of irradiated planets \citep{2005AREPS..33..493G}, CoRoT-Exo-2b appears once again to be anomalously large: its radius is about 0.3\Rjup \,larger than expected for a hydrogen-helium planet of this mass and irradiation level. However, contrary to most planets discovered to date, standard recipes to explain this large radius (heat dissipation in the interior or increased opacities) yield only a $\sim$0.15\Rjup increase, and thus are not sufficient. Larger deviations from the standard models (e.g. very large tides) or effects that may alter the radius determination (e.g. large brightness variations of the stellar surface) should be studied.

Due to its short orbital radius and the high mass, both the star and the planet exchange strong tidal forces. The Doodson constants for the star and the planet, a measure of the magnitude of the tidal forces, are of the same order of magnitude as for OGLE-TR-56b and confirms that this system is a good candidate for the study of the evolution of the system under tidal interactions. As the orbital period (1.7 days) is shorter than the stellar rotation period (4.5-5 days), the planetary rotation should be synchronized, and the orbit should decay due to tidal effects (e.g. \citealt{2002ApJ...568L.117P}). 


A thorough interpretation of the light curve, which requires a detailed modelling of the effects of the stellar activity onto both the global star luminosity and the luminosity during transits, will provide an unprecedented view of the star-planet interactions.
\begin{acknowledgements}
We thank the OHP staff for the help provided in the observations with SOPHIE. The German CoRoT team (TLS and Univ. Cologne) acknowledges the support of DLR grants 50OW0204, 50OW0603, and 50QP0701. H.J.D. and J.M.A. acknowledge support by grants ESP2004-03855-C03-03 and
ESP2007-65480-C02-02 of the Spanish Education and Science ministry.

\end{acknowledgements}
%

\end{document}